\documentclass[aps,prb,showpacs,superscriptaddress,amsmath,twocolumn]{revtex4-1}

\usepackage{amsbsy}
\usepackage{amssymb}
\usepackage{graphicx}
\usepackage{bbm}
\usepackage{bbold}
\usepackage{ulem}

\begin{document}

\title{Magnetic field oscillations of the critical current in long ballistic graphene Josephson junctions}

\author{P\'eter  Rakyta}
\affiliation{Department of Physics of Complex Systems,E{\"o}tv{\"o}s University,H-1117 Budapest, 
         P\'azm\'any P{\'e}ter s{\'e}t\'any 1/A, Hungary}

\author{Andor Korm\'anyos}
\affiliation{Department of Physics, University of Konstanz, D-78464 Konstanz, Germany}

\author{J\'ozsef Cserti}
\affiliation{ Department of Physics of Complex Systems,E{\"o}tv{\"o}s University,H-1117 Budapest, 
         P\'azm\'any P{\'e}ter s{\'e}t\'any 1/A, Hungary}

\email{rakytap@caesar.elte.hu}       
\email{andor.kormanyos@uni-konstanz.de}


\begin{abstract}
We study the Josephson current in long ballistic superconductor-monolayer graphene-superconductor 
junctions. As a first step, we have developed an efficient computational approach to 
calculate the Josephson current  in tight-binding  systems. 
This approach can be particularly useful in the long junction limit, 
which has hitherto attracted less theoretical interest but has  recently become 
experimentally relevant.  
We use this computational approach to study the dependence of the critical current 
on the junction  geometry, doping level,   and an applied perpendicular 
magnetic field $B$. In zero magnetic field we  find  a good qualitative agreement with 
the recent experiment of Ben Shalom \textit{et al.} (Reference \onlinecite{graphene_Falko}) 
for the  length dependence of the critical current. 
For highly doped samples our numerical calculations show a broad agreement with the 
results of the quasiclassical formalism. In this  case the critical current exhibits
Fraunhofer-like oscillations as a function of $B$.  
However, for lower doping levels, where the cyclotron orbit becomes comparable 
to the characteristic geometrical length scales of the system, 
deviations from the results of the quasiclassical formalism appear. 
We argue that due to the exceptional tunability and long mean free path of graphene systems 
a new regime can be explored 
where  geometrical and dynamical effects are equally important to understand
the magnetic field dependence of the critical current. 
\end{abstract}


\maketitle

\section{Introduction}

The recent progress in the fabrication techniques of graphene devices allows to obtain 
exceptionally high  mobilities with  mean free paths of several $\mu$ms in 
graphene devices~\cite{crdean2010,mayorov,crdean2013}. Thus a new field has opened for experiments, 
the electron optics of two-dimensional Dirac electrons~\cite{marcus,schonenberger2013,schonenberger2015,ozyilmaz,hu-jong}.  
Very recently, several work has made a further exciting step by contacting 
such high-quality graphene samples with superconducting 
electrodes~\cite{xudu2013,vandersypen,yacobi2015a,graphene_Falko,xudu2015,specular-andreev}
and observed a finite  Josephson current  flowing over $\mu$m distances~\cite{vandersypen,graphene_Falko}. 
In addition, the interface between the superconducting (S) and graphene (G) 
regions was found to be significantly more transparent than in previous experiments
\cite{xudu2013,morpurgo2007,lau2007,bouchiat2007,andrei,siddiqi,bouchiat2009,
leehj2011a,leehj2011b,bezryadin,schonenberger2012,gueron,leehj2013,esquinazi}. 
These experimental advances may allow to verify some of the theoretical predictions 
for graphene-superconductor  heterostructures, 
such as anharmonic phase-current relation of supercurrent at low temperatures in  
superconductor-graphene-superconductor (SGS) 
junctions in monolayer~\cite{black-schaffer2008,black-schaffer2010,hagymasi,peeters2012} 
and bilayer~\cite{peeters2012,moghaddam}
graphene, supercurrent quantization in quantum point contacts~\cite{zareyan2006,zareyan2007},
specular Andreev reflection~\cite{beenakker2006,qing-feng2009,qing-feng2011,recher}, 
detection of valley polarization \cite{beenakker2007},
interplay of strain and superconductivity~\cite{peeters2011,linder,bwang} etc.  in the near future.

The theoretical work has mainly focused on short SGS junctions to date 
\cite{moghaddam,zareyan2006,zareyan2007,linder,Carlo_short1,imura}, 
where the length of the normal region $L$ is smaller than the 
superconducting coherence length $\xi_0 = \frac{\hbar v_F}{\Delta_0}$.
In addition, it was usually assumed that the width $W$ of the junction is much larger than $L$. 
Although the long junction regime has been studied theoretically for 
superconductor-normal metal-superconductor 
(SNS) systems~\cite{ishii,bardeen,bratus1972,miller,diffusive_long_SNS,bergeret,cuevas}, 
the physics of long SGS junctions is less explored. 
An  experimental study of long ($L\gtrsim \xi_0$) and wide ($W\gg L$)
diffusive SGS junctions was presented in Reference \onlinecite{gueron}. 
However,  in recent experiments  different transport  
regimes have become accessible, where ballistic propagation  was achieved in graphene samples 
where $L\gtrsim \xi_0$ \cite{vandersypen,graphene_Falko} and/or $W/L\approx 1$\cite{vandersypen}.  
Furthermore,  the dependence of the superconducting critical 
current $I_c$ on a perpendicular magnetic field ${B}$ 
has also been measured\cite{vandersypen,yacobi2015a,graphene_Falko} in these SGS junctions. 
While References \cite{yacobi2015a,graphene_Falko} have found that the oscillations of $I_c$ 
as a function of  ${B}$  can be described, at least in  doped samples, 
by a Fraunhofer-like interference pattern, in Reference \cite{vandersypen} deviations from the 
Fraunhofer pattern have been observed for samples that are in the long junction limit 
and have an aspect ratio $W/L\approx 1$.  
Previously, deviations from the Fraunhofer-like $I_c({B})$  dependence were 
also observed in SNS junctions both in the diffusive~\cite{bouchiat2012} 
and in the quasi-ballistic limit~\cite{anomalous_osc1}, 
and the subsequent theoretical work have elucidated the 
role of the junction geometry~\cite{blatter,zagoskin1999,zagoskin2003} 
using the quasiclassical Green's function approach. It is not immediately clear, however, 
if these  theoretical results are directly applicable to SGS junctions, especially in the 
low doping regime. 

Our aim in this work is twofold. Firstly, we want to present 
a newly developed computational approach to calculate the Josephson current 
in tight-binding (TB) systems. The method is general and can be 
implemented for many TB systems, not only for graphene. 
It takes into account on equal footing the contributions coming from both the 
Andreev bound states (ABS) and the scattering states (ScS), the latter being especially  important 
in long Josephson junctions, where it is known that cancellation  
between different supercurrent contributions occur~\cite{bratus1972,affleck2013}. 
Since our method accounts for both contributions, it  can be used 
for efficient simulations of recent experimental systems~\cite{vandersypen,yacobi2015a,graphene_Falko}. 
Secondly, using the above computational method,  we study 
the length $L$ and  magnetic field dependence of the critical current in long SGS junctions. 
Although the length dependence of $I_c$ has been studied before using various 
theoretical approaches~\cite{black-schaffer2008,hagymasi,perfetto,jafari} we 
revisit this question because the recent observations in Reference~\cite{graphene_Falko}
offer the possibility to directly compare theory and experiments. 
Encouragingly, we find a good qualitative agreement between our results  and the 
experimental observations of Reference~\cite{graphene_Falko}, indicating that 
our approach  can capture important aspects of the physics of long SGS junctions. 
Regarding the magnetic field effects in long SGS junctions, to our knowledge no detailed 
study is available at present. We study the magnetic field oscillations of $I_c$ as  function
of the doping of the normal region. 
For high doping we find that the semiclassical formalism~\cite{blatter,zagoskin1999,zagoskin2003}, 
developed for ballistic Josephson junction where the normal region is a two-dimensional electron 
gas, can also describe the  oscillations $I_c$ in SGS junctions. However, for lower doping, 
where the cyclotron radius $R_c$ becomes comparable to $W$ and/or $L$, orbital effects 
can no longer be neglected and deviations from the quasiclassical results appear.

The paper is organized as follows.    
In Section \ref{sec:model} we briefly introduce the model system that we used 
in our calculations. In order to make the paper accessible to a broad audience, 
this is followed by the presentation of our main 
results for the critical current $I_c$. First, in Section \ref{sec:zero-field results}
we discuss the length dependence of $I_c$ and also the current-phase relation. 
The effect of the magnetic field on $I_c$ is treated in Section \ref{sec:magnetic-field-osc}. 
A general numerical approach to calculate the Josephson current for TB Hamiltonians is 
presented in Section \ref{sec:TB-method}, while some of the relevant details of the TB 
model used in this work is given in Section \ref{sec:TB-details}. 
Finally, we conclude with Section \ref{sec:summary}.

\section{The model}
\label{sec:model}

We first briefly describe the model we employed to calculate the Josephson current in SGS junctions,
further details can be found in Section \ref{sec:TB-details}. 
In the normal conducting region of length $L$  and width $W$ we use 
the nearest-neighbour TB model of graphene 
\cite{wakabayashi,graphene-review} with the Hamiltonian 
\begin{equation}
 \hat{H} = \sum_{i}\varepsilon_i c_{i}^{\dagger} c_i - \sum_{\langle ij\rangle}\gamma_{ij}c_{i}^{\dagger} c_j+ h.c.
\end{equation}
Here $\varepsilon_i$ is the on-site energy on the atomic site $i$,  $\gamma_{ij}$ 
is the hopping amplitude between the nearest neighbour atomic sites $\langle ij\rangle$ 
in the graphene lattice, and $c_{i}^{\dagger}$ ($c_i$) creates (annihilates) an electron at site $i$. 
The magnetic field can be incorporated  by means of the Peierls substitution \cite{peierls}:
\begin{equation}
 \gamma_{ij} = \gamma\;{\rm Exp}\left( \frac{2\pi{\rm i}}{\phi_0} \int\limits_{\mathbf{R}_j}^{\mathbf{R}_i} 
 \mathbf{A(\mathbf{r})}{\rm d}\mathbf{r} \right) ,
 \label{eq:TB-peierls}
\end{equation}
where $\phi_0 = h/e$  is the flux quantum,  $\mathbf{A}(\mathbf{r})$ denotes 
the vector potential and the vector $\mathbf{R}_i$  points to the $i$th atomic site in the lattice.
The spatial dependence of $\mathbf{A}(\mathbf{r})$ is such that it yields a homogeneous perpendicular magnetic 
field $\mathbf{B} = (0,0,B_z)^T$  in the normal region and zero field in the superconducting regions, 
see Section \ref{subsec:Bfield}.  

The superconducting regions are modelled by a highly doped graphene region\cite{Carlo_short1}  
of width $W$ and {open boundary condition in the transport direction.}
It is assumed that a finite on-site pair-potential $\Delta_0 e^{i\varphi_{L,R}}$ is induced 
by proximity effect in the left (L) and right (R) electrodes. 
We note that our methodology would allow for other models of the superconducting regions 
as well \cite{martin-rodero}. 
For the superconducting pair-potential we assume a  step-like change at the 
normal-superconductor (NS) interfaces:
\begin{equation}
 \Delta(x) = \left\{ \begin{array}{ll}
		      \Delta_0 e^{{\rm i}\varphi_L} & \textrm{if } x< 0 \\
                      0 & \textrm{if } 0\leq x\leq L \\
		      \Delta_0 e^{{\rm i}\varphi_R} & \textrm{if } x>L
                     \end{array}
\right..
\end{equation}
Here $\varphi_L$ ($\varphi_R$) is the phase of the pair-potential in the left (right) lead 
and we will denote by $\delta\varphi=\varphi_R-\varphi_L$ the phase difference. 
Our main interest in this work is to study SGS junctions where there is a significant 
difference between the doping levels of the S and N regions:  
$\lambda_S \ll \lambda_N$ where $\lambda_{S(N)}$ is the Fermi wavelength in the 
superconductor (normal) region. Moreover, 
the junction is long $L/\xi_0\gtrsim 1$ with respect to the coherence length $\xi_0$.
In this case we expect that the detailed spatial dependence of $\Delta_0$ in the vicinity
of the normal-superconductor interface is not very important and therefore the above 
approximation should give qualitatively correct results. 
{Indeed, References \cite{black-schaffer2008,black-schaffer2010,peeters2012,peeters2013,alidoust} 
have shown that the self-consistent calculation of $\Delta(x)$ in clean SGS junctions
is most important for (i) short junctions, (ii) no Fermi-level
mismatch between the S and N regions, (iii)  temperatures close to  $T_c$.}

The simulation of realistic samples on the micrometer length scale is quite challenging in the TB 
framework due to the huge number of the atomic sites. Part of the problem can be circumvent by using 
an efficient numerical approach, see Section \ref{subsec:Heff} for details. Moreover, 
we expect that experimentally relevant informations can be extracted from TB systems 
that follow certain scaling laws but imply significantly lower computational costs. 
Such an approach has proved to be very useful recently in the calculation of 
normal transport \cite{graphene_antidot_sajat,TB_scaling} for mesoscopic graphene
structures. We expect that as long as  the characteristic dimensions $W$  and $L$ of 
the system are much larger than the lattice constant of graphene, 
the {same} physical behavior should be observed in systems with the 
same $L/\xi_0$, $W/L$, $\frac{2W}{\lambda_N}$, {$\frac{\xi_0}{\lambda_N}$, $\frac{T}{T_c}$} 
and $L/l_B$ control parameters, where
$\lambda_N=\frac{E_F}{\hbar v_F}$ is the Fermi wave number in the normal 
region $l_B=\frac{\hbar}{|eB|}$ is the magnetic length 
{and $T_c\approx  \Delta_0/(1.76 k_B)$ is the critical temperature}.
{ We are interested here in the bulk properties of the supercurrent, i.e., 
we need to ensure that edge effects do not play role. In most of our calculations we used zigzag 
nanoribbons,} { however, we have checked that we obtain very similar results 
for armchair nanoribbon as well. 
(see Section \ref{sec:zero-field results}). Therefore we expect 
that our results would not change for more general edges either.}


\section{Zero magnetic field results}
\label{sec:zero-field results}



 \subsection{Length dependence of $I_c$}
 \label{subsec:length-dependence}

An  important property of long Josephson junctions is the dependence of the critical current $I_c$ 
on the junction length $L$, which was measured recently in Reference \cite{graphene_Falko}. 
In general, at zero temperature $I_c$ is given by the relation\cite{Carlo_revmod}
\begin{equation}
 I_c = \alpha \frac{e|\Delta_0|}{\hbar}N, 
 \label{eq:alpha}
\end{equation}
where $N$ is the number of open channels in the normal region:
$
 N = \frac{E_FW}{\pi\hbar v_F} = \frac{Wk_F}{\pi} = \frac{2W}{\lambda_F}.
 $
The  dimensionless coefficient $\alpha$ can depend on a number of factors, 
such as the junction transparency, the presence of a $p-n$ junction or other disorder 
induced by the superconducting contacts, the doping level in the normal region \cite{Carlo_short1} etc. 
However, for the  case considered here (no disorder at the SG interface)  and in the long 
junction limit $\alpha$ is expected to be a function of the ratio $L/\xi_0$ only \cite{bardeen,bratus1972}.
\begin{figure}
\centering
 \includegraphics[scale=0.45]{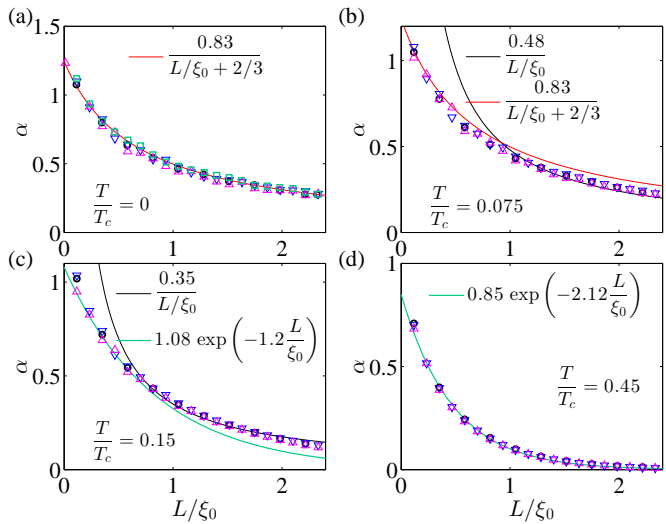}
 \caption{Length dependence  of $I_c$ for different temperatures $T/T_c$.
 Different symbols correspond to calculations performed with {zigzag} 
 ribbons of different width: 
 $W=149\, {\rm r}_{cc}$ ($\vartriangle$), $W=299\, {\rm r}_{cc}$ ($\triangledown$), and 
 $W=449\, {\rm r}_{cc} $ ($\circ$), where  $ {\rm r}_{cc}$ is the carbon-carbon bond length. 
 { The calculations shown with $\square$ were obtained for an armchair nanoribbon 
  of width $W=301\, {\rm r}_{cc}$. Solid lines show the results of fitting, the obtained fit 
  parameters are indicated in each figures.}
 The chemical potential in graphene is $\mu_N=80\Delta$.
 \label{fig:alpha}} 
\end{figure}

{ First, in Figure \ref{fig:alpha}(a) we present the zero temperature  calculations for $\alpha$.
The good agreement between the calculations for different widths { and edge types} 
indicates  that these results are free of finite size { and edge} effects. 
One can see that the numerical results can be fitted with a function $A/(L/\xi_0+C)$, where 
$A$ and $C$ are fitting parameters.
}
For $L/\xi_0 \gg C$ the critical current falls off as $1/L$, whereas 
in the short junction limit $L/\xi_0\ll 1$ it goes to a finite value and reproduce the analytical 
prediction of Reference~\onlinecite{Carlo_short1}:  for 
$L/\xi_0\rightarrow 0$  and $\lambda_S \ll \lambda_N$  the value of $\alpha$ is $\approx1.22$. 
Next, in Figures \ref{fig:alpha}(b)-(d) we show the $I_c(L)$ curves for low, but finite temperatures $T$. 
As the temperature is increased from $T=0$,  both fit parameters $A$ and $C$ decrease and 
eventually  there is a  temperature range where the parameter $C$ becomes very small 
so that  $I_c$ falls off as $\sim 1/L$ for $L/\xi_0\gtrsim 1.0$ [Figures \ref{fig:alpha}(b)-(c)]. 
Such a $\sim 1/L$  dependence of $I_c$ was recently observed in Reference~\cite{graphene_Falko}.  
It was suggested that this peculiar $I_c(L)$ dependence is a 
signature of the  SGS junctions  being truly ballistic and in the long junction limit. 
Our calculations support this conclusion, nevertheless, it would be interesting to measure 
$I_c(L)$ at lower temperatures in order to map out experimentally 
the $T$ dependence  of the parameters $A(T)$ and $C(T)$ and compare it to our results.   
It is important to note the following: 
(i) Taking the  $T_c$ of bulk niobium, the experimental data of Reference~\onlinecite{graphene_Falko} 
correspond to $T/T_c=0.045$, 
i.e., to lower temperature than shown in Figure \ref{fig:alpha}(b). 
However, if one assumes that the induced pair potential in the graphene is smaller 
than the bulk value of $\Delta$, i.e., it corresponds to a smaller $T_c$, then 
the agreement with our calculations becomes better. (ii) The   values of $\alpha$  in  
Figure \ref{fig:alpha}(c) significantly overestimate the experimental ones  reported in  
Reference~\onlinecite{graphene_Falko}. 
{We think this is due to the fact that we have assumed perfectly transparent 
SG interface and no disorder in the normal part of the junction, while
in   Reference~\onlinecite{graphene_Falko} the SG interface had a finite transparency and  
a p-n junction was probably formed due to the doping effect of the contacts. 
We leave the study of these effects for a future work.} 
Finally, as shown in Figures \ref{fig:alpha}(d), as the temperature is further increased such that 
the energy scale $k_B T$ becomes non-negligible with respect to the 
ballistic Thouless energy $\frac{\hbar v_{F}}{L}$ \cite{bardeen,miller}, 
$I_c$ becomes exponentially suppressed: $I_c\sim e^{-c L/\xi_0}$, where $c$ is a fitting parameter.

\subsection{Current-phase relation}
 \label{subsec:current-phase}

\begin{figure}[htb]
\centering
 \includegraphics[scale=0.42]{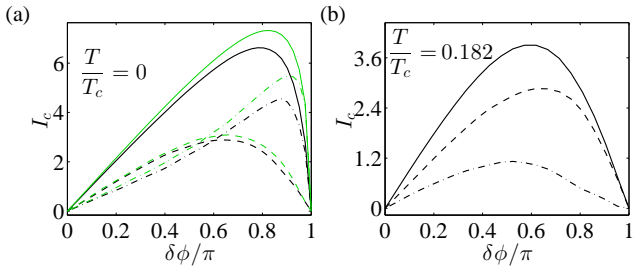}
 \caption{The Josephson current  as a function of the phase difference 
 $\delta\phi=\phi_R-\phi_L$ between the superconducting electrodes 
 (in units of $e\Delta_0/\hbar$). 
 The parameters are  $\mu_N=32\Delta_0$, $W/L=1.15$, and $L/\xi_0=1.48$. 
 The total current is denoted by solid lines, and the separate contributions of the 
 scattering states (dashed-dotted) and  of the Andreev bound states (dash-dotted) 
 are also shown.  Black lines show the results obtained using zigzag nanoribbons,
 green lines indicate armchair results. {The ribbon width was $W=300\, {\rm r}_{cc}$.}
\label{fig:Josephson_TB}}
\end{figure}

Regarding  the current-phase relation (CPR), it has long been known that the  CPR in long SNS 
Josephson junctions~\cite{ishii,bardeen,bratus1972} is substantially different from the 
$\propto \sin(\delta\phi)$ relation valid in the short junction limit~\cite{Kulik_shortSNS}.
In the one-dimensional TB model  studied in Reference~\onlinecite{cini} the deviation 
from the $\propto \sin(\delta\phi)$ dependence was explained by the 
contribution of the scattering states to the total current which becomes 
comparable to the contribution of the Andreev bound states  as the length of the 
junction increases. A partial cancellation effect between the different contributions to 
the Josephson current in SNS junctions was also pointed out by References~\onlinecite{bardeen,bratus1972}. 
Recently,  anharmonic CPR was found in the calculations of  
References~\cite{black-schaffer2008,black-schaffer2010,peeters2012}  for SGS junctions. 
We have also calculated the CPR using our method for a long SGS junction. 
As one can see in Figure \ref{fig:Josephson_TB}(a), at zero temperature the  CPR deviates significantly 
from the $\propto \sin(\delta\phi)$ dependence and for $\delta\phi/\pi \lesssim 0.3$ it is a linear 
function of $\delta\phi/\pi$. 
{The contribution coming from the ScS (dashed-dotted line) is of the same 
magnitude as the contribution of the ABSs (dashed line).} 
In Figure \ref{fig:Josephson_TB}(a)  the 
two contributions have the same sign, however, this is not always the case: 
 calculations not shown here indicate that depending on  the $L/\xi_0$ ratio 
the supercurrent due to the ScS can be either positive or negative.
For finite temperatures [see Figure \ref{fig:Josephson_TB}(b)], 
similarly to  SNS junctions \cite{bardeen},  the CPR  acquires  a harmonic dependence 
on $\delta\phi$. {In these wide ribbons the major characteristics of the CPR 
do not depend on whether zigzag or armchair nanoribbons are used in the calculations.}

To briefly summarize the  results presented in this section, we find that 
 our results show a good  qualitative agreement with 
i) the measurements of Reference~\onlinecite{graphene_Falko} for the length dependence of $I_c$, 
ii)  with previous theoretical works~\cite{black-schaffer2010,peeters2012} for the CPR, even though 
we do not calculate the pair potential $\Delta$  self-consistently.
This   suggests  that in long SGS JJs, if there is  a  finite Fermi-level
mismatch at the SG  interface,  the  self-consistent calculation of $\Delta$ is less 
important than in the short junction limit with no Fermi-level mismatch.


\section{Oscillations of the critical current in perpendicular magnetic field}
\label{sec:magnetic-field-osc}

We now turn to the properties  of the critical current 
in the presence of a perpendicular magnetic field $B$. Oscillations of
$I_c$ have been measured in  several recent 
experiments~\cite{vandersypen,yacobi2015a,graphene_Falko} but have 
not yet received much theoretical attention. 
It is well known that in tunnel junctions $I_c$ exhibits 
Fraunhofer-like oscillations as a function of the piercing 
magnetic flux $\Phi$ with a period 
$\phi_0=h/2e$ and oscillation minima at integer multiples 
of the flux quantum $\phi_0$:
$
 I_c(B) = I_c(0) 
 \left|
 \frac{{\rm sin}(\pi\Phi/\phi_0)}{\pi\Phi/\phi_0}
 \right|, 
$
where $I_c(0)$ is the zero-field critical current. 
The long junction limit in SNS systems was studied in 
References~\cite{blatter,zagoskin1999,zagoskin2003,svidzinskii} using the 
quasiclassical Green's function formalism.  It was pointed out that 
the magnetic field oscillation of $I_c$ also depend on the 
geometry of the junctions. For wide and long ballistic junctions,  
where $W\gg L\gg \xi_0$,   the critical current oscillates as~\cite{zagoskin1999,svidzinskii}
\begin{equation}
 I_c(B)=I_c(0)\frac{(1-\{\Phi/\phi_0\}) \{\Phi/\phi_0 \}}{|\Phi/\phi_0|}, 
\label{eq:Ic-wide}
\end{equation}
 $\{x\}$ denoting the fractional part of $x$.  
The oscillation pattern given by Eq.~(\ref{eq:Ic-wide}) is very similar to 
the Fraunhofer-like oscillations in tunnel junctions, except for $\Phi/\phi_0\ll 1$. 
However, deviations from  Eq.~(\ref{eq:Ic-wide}) were observed 
in the measurements of Reference~\onlinecite{anomalous_osc1}  in  quasi-ballistic SNS junctions. 
Subsequent theoretical work \cite{blatter,zagoskin1999}  showed 
that geometrical effects become important when $W\sim L\gg \xi_0$. 
In particular, Reference~\onlinecite{blatter} found that the periodicity of $I_c$  
as a function of magnetic field changes from $\phi_0$ [see Eq.~\ref{eq:Ic-wide}]
to $2\phi_0$ as the flux through the junction increases and at low temperatures 
the crossover to the $2\phi_0$ periodicity appears  at a flux $\sim \phi_0 W/L$. 
Regarding the recent experiments in SGS junctions,   
Fraunhofer-like pattern  for $I_c(B)$ was found  in Reference~\onlinecite{graphene_Falko} for 
wide ($W\gg L$) junctions, whereas  Reference~\onlinecite{vandersypen} measured periodicity that was  
longer than  $\phi_0$ for junctions with aspect ratios $W/L\approx 1$.

An important assumption behind the quasiclassical 
formalism \cite{blatter,zagoskin1999,zagoskin2003,svidzinskii}
is that the effect of magnetic field can be taken into account through a phase factor 
that the wave function of the quasiparticles acquires  
along  classical trajectories that are  straight lines.  
This is a good approximation when the cyclotron radius of the particles is much 
larger than other characteristic length scales in the system.  

However, compared to  traditional SNS systems, the exceptional tunability 
of the  state-of-the-art 
graphene systems, combined with the use of Nb~\cite{graphene_Falko} 
or MoRe~\cite{vandersypen} superconductors allows, in principle, to reach a regime 
where the  size of the cyclotron radius $R_c$ is 
comparable to  $W$ and/or $L$ already for relatively low magnetic fields such that the 
effect of magnetic field on  the superconducting electrodes  can still be neglected. 
Considering the semiclassical cyclotron radius $R_c$ in graphene
$ R_c = \frac{E_F l_B^2}{\hbar v_F}$
one can see that for a charge density $n_e\approx 10^{12}\frac{1}{cm^2}$ 
the cyclotron radius is  $\approx 1\mu$m  for a $B\approx 0.1 T$. 
Thus $R_c$ becomes an important length scale when the chemical 
potential approaches the charge neutrality point.
This regime is expected to harbour  rich physics, because geometrical effects,  
discussed above, and dynamical effects related to the 
curved semiclassical trajectories  of the quasiparticles  are equally important. 
It is then not clear if the quasiclassical 
formalism \cite{blatter,zagoskin1999,svidzinskii}, which neglects the dynamical effects,
can still give a good description. The possible importance of this 
regime has also been discussed in  Reference~\onlinecite{graphene_Falko}. 

In the following we take  the chemical potential $\mu_N$ in the normal region as 
a tuning parameter discuss separately the \textit{high doping limit}, when 
$R_c\gg L, W$ holds for the considered magnetic field range, and the 
\textit{low doping range}, where $R_c\gtrsim L, W$ can be reached for 
relatively small magnetic fields. 
In all our calculations we assume that the  effect of magnetic field on 
the pair potential  in the superconducting contacts can be neglected.

\subsection{High doping limit}
\label{subsec:highdoping}

\begin{figure}
\centering
 \includegraphics[width=8.5cm]{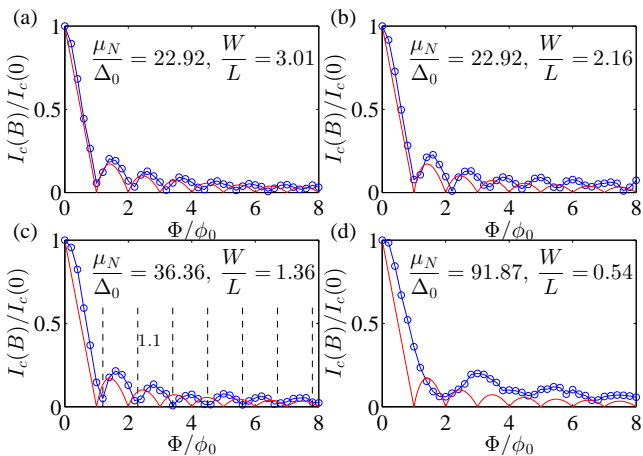}
 \caption{The normalized critical current as a function of the magnetic flux piercing  
  the  SGS junction (circles). In all figures $L/\xi_0=1.63$, and  the temperature was  $T/T_c=0.013$.
  In (c) the vertical dashed lines show the position of the current minima. The distance between 
  the minima is $1.1\phi_0$ in the plotted magnetic field range. 
The solid (red) line represents the standard Fraunhofer-like oscillation pattern given 
by  Eq.~(\ref{eq:Ic-wide}). 
\label{fig:fraunhofer}} 
\end{figure}

To see if the quasiclassical theory~\cite{blatter,zagoskin1999} also applies 
to SGS junctions when the normal region is strongly doped, 
we calculate $I_c(B)$ for different $W/L$ ratios. 
We have found that the oscillations depend only weakly 
on the exact value of $L/\xi_0$ and on the temperature (in the  low temperature limit  
relevant for recent experiments~\cite{vandersypen,yacobi2015a,graphene_Falko}). 
Thus, we present our results only for  particular temperature and $L/\xi_0$ values.
Figure~\ref{fig:fraunhofer} shows the magnetic 
oscillations calculated for different aspect ratios $W/L$ \cite{calcs-detail}. 
{ These results were obtained for zigzag nanoribbons but 
we have checked that the results do not change qualitatively for armchair nanoribbons.}
For wide junctions [Figure ~\ref{fig:fraunhofer}(a)], we recover the Fraunhofer-like pattern 
of the oscillations with minimums at integer multiples of the flux quantum $\phi_0$, 
see Eq.~(\ref{eq:Ic-wide}). Deviations from  the ideal curve Eq.~(\ref{eq:Ic-wide}) 
start to appear only for  magnetic fluxes $\Phi\gtrsim 4\phi_0$.
As the aspect ratio $W/L$ decreases [Figures \ref{fig:fraunhofer}(b)-(c)], 
the periodicity of $I_c$ becomes longer than $\phi_0$, even for smaller magnetic fields. 
Interestingly, for $W/L\sim 1$ [Figure \ref{fig:fraunhofer}(c)] the oscillation period 
is roughly constant in the considered magnetic field range. 
Finally, in Figure \ref{fig:fraunhofer}(d) one can see that in narrow samples the 
first minimum in $I_c$ is at $2\phi_0$ and the current does not go to zero.   
These results are in broad agreement with  the quasiclassical calculations of 
Reference~\onlinecite{blatter}  indicating that $I_c(B)$ 
in highly doped graphene samples can essentially be described by the theory used
previously for SNS junctions in References~\onlinecite{blatter,zagoskin1999,svidzinskii}. 
As mentioned above, deviations from the Fraunhofer-like pattern, similar to the ones shown 
in Figures~\ref{fig:fraunhofer}(c) and (d) have been measured 
recently in Reference~\onlinecite{vandersypen} for samples with $W/L\approx 1$. 
However, a  more quantitative comparison of our calculations with   Reference~\onlinecite{vandersypen} 
is difficult because i) due to the inevitable disorder
at the edges, the effective width of the samples may be smaller than the geometrical width, and 
ii) the current oscillations are likely to depend on the properties of the $p$-$n$ junction
formed at the SG interface which is not taken into account in our calculations.

\subsection{Low doping limit}
\label{subsec:lowdoping}

\begin{figure}
\centering
 \includegraphics[width=8.5cm]{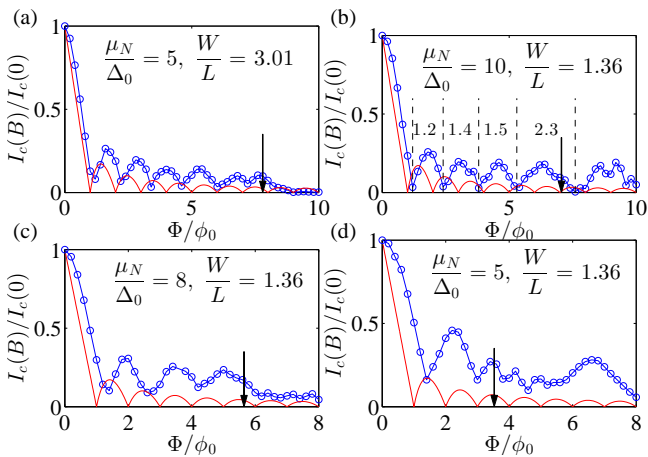}
 \caption{The normalized critical current as a function of the magnetic flux piercing the 
 SGS junction. 
 The parameters used in these calculations were $L/\xi_0=1.63$, and  $T/T_c=0.013$.
 The solid (red) line represents the standard Fraunhofer-like oscillation pattern 
 given by Eq.~(\ref{eq:Ic-wide}).
 The dashed vertical lines in panel (b) indicate the minima of the current. 
 Vertical arrows indicate  the $\Phi/\phi_0$ values 
 where the cyclotron radius equals the length of the junction, i.e., $R_c=L$. 
 \label{fig:fraunhofer_EF}} 
\end{figure}

We now consider the case when the doping of the normal region approaches the 
charge neutrality point, but it is still large enough so that
the formation electron-hole puddles can be neglected. 
The most interesting results in this  parameter range are shown in Figure~\ref{fig:fraunhofer_EF}.  
As in the high doping case, the current oscillations depend on the aspect ratio $W/L$. 
For wide junctions, as in Figure~\ref{fig:fraunhofer_EF}(a), the oscillations are similar 
to the  strong doping case shown in Figure~\ref{fig:fraunhofer}(a) but the oscillation 
period is somewhat longer and the amplitude, relative to $I_c(0)$,  is larger. 
For $L> R_c$, however, the current is strongly suppressed. 
In Figures~\ref{fig:fraunhofer_EF}(b)-(d) we  calculate the current for a narrower junction 
as the doping is decreased.  Comparing Figure~\ref{fig:fraunhofer_EF}(b) and 
Figure~\ref{fig:fraunhofer}(c), where the same  geometrical parameters were used, one can 
see that for smaller doping  dynamical effects  influence the period of the current oscillations.
(As indicated in Figure~\ref{fig:fraunhofer_EF}(b), the  ratio  $L/R_c\approx 1 $ 
is obtained for $\Phi/\phi_0\approx 7.8$ in this case. )
While in the strongly doped case [Figure \ref{fig:fraunhofer}(c)] the period of oscillations
is $\approx 1.1\phi_0$ in the shown magnetic field range,  for smaller doping the distance 
between consecutive minima grows with the magnetic field.  
As the doping level is further reduced the oscillations of $I_c$ become rather complex  
for magnetic fields where $L> R_c$  is fulfilled, see Figures~\ref{fig:fraunhofer_EF}(c) and (d).  
The regime $\Phi/\phi_0 \gtrsim 5.8$ [$\Phi/\phi_0 \gtrsim 3.8$] in 
Figure~\ref{fig:fraunhofer_EF}(c) [Figure~\ref{fig:fraunhofer_EF}(d)] illustrates that 
geometrical effects (due to $W\sim L$) and dynamical effects ($R_c\sim L$)
can both strongly affect the oscillation pattern. 
Note,  that for the parameters used in  
Figure~\ref{fig:fraunhofer_EF}(b)-(d) the diameter of the cyclotron motion is still 
smaller than the  geometrical parameters, i.e.,
$2 R_c > W, L$ and therefore no quantum Hall edge states are formed.

The regime $R_c\gtrsim L$ has recently been considered in Reference~\onlinecite{graphene_Falko}, 
where  the sample dimensions were $W\gg L\gtrsim \xi_0$.  
The suppression of the supercurrent was  discussed  in terms
of the constraints that classical trajectories, corresponding to electron-hole pairs, 
have to fulfill in order that Cooper-pairs can be transferred between the superconductors. 
It was argued, amongst other,  that electron-hole trajectories should not drift 
farther away than the size of  Cooper-pair wave packet in the superconductor. 
For wide junctions [Figure~\ref{fig:fraunhofer_EF}(a)], our results seem to be in 
qualitative agreement with the semiclassical picture put forward in Reference~\onlinecite{graphene_Falko}. 
For junctions where $W \sim L \sim \xi_0$ [Figures~\ref{fig:fraunhofer_EF}(b)-(d)], the situation is 
somewhat different from the wide junction case  because here the electron-hole trajectories cannot 
drift away to large distances and are more likely to form (nearly) closed orbits and hence 
bound states. This may explain why the supercurrent is not suppressed.


\section{Tight-binding approach to calculate the DC Josephson current}
\label{sec:TB-method}

In this section we describe the  TB approach we used to calculate the DC Josephson current. 
It is a generalization of the approach developed to calculate two-terminal normal transport 
\cite{sanvito1999,smeagol-paper} to the case when  both terminals are superconducting and the 
main interest is not the scattering matrix (and the differential conductance) but the 
current flowing due to the superconducting phase  difference between the terminals. 
As already mentioned, the  method is general and can be implemented  for many TB system.
It is a generalization of the one-dimensional TB work by References~\onlinecite{cini,cbena} 
to nanoribbon geometries 
and 
would also allow for extension  to multi-terminal systems 
studied, e.g., in  References~\onlinecite{qing-feng2009,qing-feng2011,akhmerov2014}. 
The actual calculations presented in Sections \ref{sec:zero-field results} and 
\ref{sec:magnetic-field-osc} 
were performed with the \texttt{EQuUs} software \cite{equus}. 


\subsection{The general setup}
\label{subsec:general-setup}

\begin{figure}[htb]
\centering
\includegraphics[width=7cm]{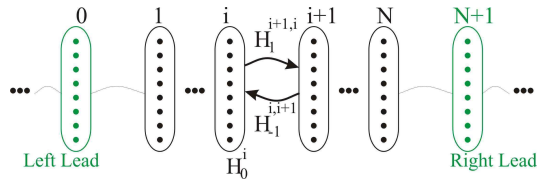}
 \caption{The atomic sites are grouped into a sequence of slabs that are connected by nearest neighbour couplings. 
  The central region is formed by the slabs from $1$ to $N$, while the slabs $0$ and $N+1$ 
 are the surface slabs of the left and right electrodes connected to the normal region.
 \label{fig:TB_system}}
\end{figure}

The studied system, including the central region and the electrodes, is schematically depicted in 
Fig.~\ref{fig:TB_system}.
The atomic sites of the system are arranged into slabs that are coupled to each other by nearest 
neighbour coupling matrices $H_1^{i,i+1}$ and $H_{-1}^{i+1,i}$.
The $i$th slab contains of $N_i$ sites and is described by a Hamiltonian $H_0^i$.
The central  (scattering) region is formed by the slabs from $1$ to $N$, 
while the slabs $0$ and $N+1$ are the surface slabs of the infinite left and right (superconducting) 
electrodes. 
Thus, the Hamiltonian of the infinite system can be organized in the following block-diagonal form:
\begin{equation}
 \hat{H} = \left( 
           \begin{array}{cccccc}
	  \dots & \dots & \dots & \dots & \dots & \dots \\
	  \dots & H_0^{i-1} & H_1^{i-1,i} & 0 & 0 & \dots \\
	  \dots & H_{-1}^{i,i-1} & H_0^{i} & H_1^{i,i+1} & 0 & \dots\\
	  \dots & 0 & H_{-1}^{i+1,i} & H_0^{i+1} & H_1^{i+1,i+2} & \dots  \\
	  \dots & 0 & 0 & H_{-1}^{i+2,i+1} & H_0^{i+2} & \dots		\\
	  \dots & \dots & \dots & \dots & \dots & \dots
	  \end{array}
	  \right)
	   \label{eq:fullHam}
\end{equation}
Let us label the eigenvalues and eigenstates of the Hamiltonian (\ref{eq:fullHam}) by $E_n$ and $\Psi_n$, 
respectively.
The index $n$ labels both the bound and scattering states that are formed in the system. 
In the latter case $n=(m,k)$ stands for a pair made of the discrete transverse quantum number $m$ and the 
wave vector $k$ describing  a propagating incoming state in one of the leads.
Generally, the Green's function of the studied system can be written as:
\begin{equation}
 G(z) = \sum\limits_n\frac{|\Psi_n\rangle\langle\Psi_n|}{z-E_n} \label{eq:G}\;.
\end{equation}
The normalization of the eigenstates $\Psi_n$ in Eq.~(\ref{eq:G}) is straightforward for the bound states, 
namely $\langle\Psi_{n_1}|\Psi_{n_2}\rangle=\delta_{n_1n_2}$.
The scattering states, on the other hand, are normalized to unit incoming current \cite{sanvito1999, sanvito2008}.


\subsection{The expectation value of the current operator}
\label{subsec:expct-current-op}

Due to current conservation, the charge current is equal between any pair of 
slabs of the studied system. Therefore, as we will show, it is sufficient to calculate 
the Green's function only on a  couple of  slabs,  leading to a numerical efficient method.  
To this end let us now introduce the operator $P^i$ projecting on the subspace of 
the $i$th slab of the system.
The matrix form of the projector $P^i$ reads
\begin{equation}
 P^i=    \left(
          \begin{array}{cccccc}
	  \dots & \dots & \dots & \dots & \dots & \dots \\
	  \dots & 0 & 0 & 0 & 0 & \dots \\
	  \dots & 0 & I_{N_i} & 0 & 0 & \dots\\
	  \dots & 0 & 0 & 0 & 0 & \dots  \\
	  \dots & 0 & 0 & 0 & 0 & \dots		\\
	  \dots & \dots & \dots & \dots & \dots & \dots
	  \end{array}
	  \right)\;, 
	  \label{eq:matrix_Pi}
\end{equation}
where $I_{N_i}$ is an $N_i\times N_i$ unity matrix.
One can notice that the projectors $P^i=\left(P^i\right)^{\dagger}$ are hermitian operators.
In addition, the operator $P^{\{i\}}$ projecting on a set of slabs $(i_1,i_2,\dots,i_p)$ can be 
given as a sum of the individual projectors
\begin{equation}
 P = \sum\limits_{q=1}^p P^{i_q}\;.
\end{equation}
Thus, the projected Green's function on the slabs $(i_1,i_2,\dots,i_p)$ can be given as $G^P(z) =  PG(z)P^{\dagger}$.
Using the Green's function given in Eq. (\ref{eq:G}), one would obtain the following projected Green's function
\begin{equation}
 G^P(z) = \sum\limits_n\frac{P|\Psi_n\rangle\langle\Psi_n|P^{\dagger}}{z-E_n} = 
 \sum\limits_n\frac{\Omega_n}{z-E_n}|\Phi_n\rangle\langle\Phi_n|\;, \label{eq:GP}
\end{equation}
where
\begin{equation}
 \Phi_n = \frac{P\Psi_n}{\sqrt{\Omega_n}}
\end{equation}
are the projected wave functions normalized to unity by the factor $\Omega_n = \langle\Psi_n|P|\Psi_n\rangle$.

\begin{figure}[htb]
\centering
\includegraphics[width=7cm]{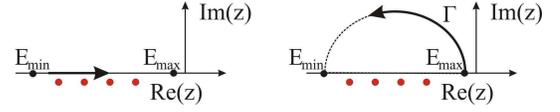}
 \caption{Since the retarded Green's function is analytical in the upper half of the complex plane, the energy 
 integral on the real axis equals to the integral along the contour $\Gamma$. The red dots below the real axis 
 represents the singularities of the retarded Green's function corresponding to the discrete bound states.
 \label{fig:Gamma_path}}
\end{figure}

The expectation value of the current operator $\hat{I}$ with respect to a 
projected state $\Phi_n$ can be calculated as follows:
\begin{eqnarray}
 \langle \hat{I}\rangle_n &= \langle\Phi_n|\hat{I}|\Phi_n\rangle = {\rm Tr}\left( \hat{I}|\Phi_n\rangle\langle\Phi_n|\right) \nonumber\\
 &= \int\limits_{E_{\rm min}}^{E_{\rm max}} dE\; \delta(E-E_n) {\rm Tr}\left( \hat{I}|\Phi_n\rangle\langle\Phi_n|\right)= \nonumber\\
  &= -\frac{1}{\pi} {\rm Im} \lim\limits_{\eta\rightarrow 0^+}\int\limits_{E_{min}}^{E_{max}}dE\; 
  \frac{{\rm Tr}\left( \hat{I}|\Phi_n\rangle\langle\Phi_n|\right)}{E-E_n + i\eta}
\label{eq:expA}
\end{eqnarray}
Here $E_{\rm min}$ and $E_{\rm max}$ are  arbitrary energies such that $E_{\rm max}-E_{\rm min}\le BW$, 
where $BW$ is the bandwidth of the system.
To obtain the total current $I_{c,tot}$, we  will need to sum over all projected states $\Phi_n$.
Using the projected Green's function $G^P$ introduced in Eq.~(\ref{eq:GP}),   
one finds that  
\begin{equation}
 \sum\limits_{E_{min}<E_n<E_{max}}\langle \hat{I}\rangle_n = -\frac{1}{\pi} 
 {\rm Im} \int\limits_{E_{min}}^{E_{max}}dE\; {\rm Tr}\left( G^P(E^+)\hat{I}\right), 
 \label{eq:sumA}
\end{equation}
where $G^P(E^+)=\lim\limits_{\eta\rightarrow0^+} G^P(E+{\rm i}\eta)$ is the retarded Green's function.
We remind, that the retarded Green's function may have poles corresponding to bound states which 
would lead to complications in the  numerical evaluation of Eq.~(\ref{eq:sumA}). 
Therefore, making use of the fact that retarded Green's function is 
analytical in the upper half-plane (${\rm Im}(z)>0$) of the complex energy, 
the integral can be performed along a path $\Gamma$ in the complex plane as shown in Fig. \ref{fig:Gamma_path}.
Thus, the sum of the expectation values of the current operator between energies $E_{\rm min}$ and $E_{\rm max}$ 
can be calculated by the formula
\begin{equation}
 \sum\limits_{E_{min}<E_n<E_{max}}\langle \hat{I}\rangle_n = \frac{1}{\pi} 
 {\rm Im} \int\limits_{\Gamma}dz\; {\rm Tr}\left( G^P(z)\hat{I}\right)\;. 
 \label{eq:sumA_contour}
\end{equation}
The total current $I_{c, tot}$ would also depend on the thermal occupation of the electronic states. 
If we have normal conducting terminals then  $I_{c, tot}$ can be expressed as 
\begin{equation}
 I_{c,tot}=\sum\limits_{E_{min}<E_n<E_{max}}\langle \hat{I}\rangle_{n,T} = \frac{1}{\pi} 
 {\rm Im} \int\limits_{\Gamma}dz\; f(z){\rm Tr}\left( G^P(z)\hat{I}\right). 
 \label{eq:sumA_contour_fermi}
\end{equation}
where  $f(z)$ is the Fermi distribution function. 
In the case of superconducting terminal the situation is somewhat more complicated but 
we leave the discussion of superconductor-normal-superconductor systems to Section \ref{subsec:sns-system}.  
Here we only note that for the evaluation of Eq.~(\ref{eq:sumA_contour}) one does not need to know explicitly 
the specific energies $E_n$ of the bound states.  Choosing $E_{\rm min}$ and $E_{\rm max}$
appropriately the contributions of both the  scattering and bound states are automatically taken into account. 
However, one has to choose such a contour $\Gamma$ which does not enclose ``unwanted poles'', e.g., 
in the case of Eq.~(\ref{eq:sumA_contour_fermi} it 
avoids the poles of $f(z)$ located at the energies $Z_l = {\rm i}(2l+1)\pi k_BT$.

As mentioned, due to current conservation it is sufficient to calculate  $I_{c,tot}$  
between two arbitrary slabs, and for practical reasons we choose the surface slabs of 
the central region (see the $1$st and the $N$th 
slabs in Fig. \ref{fig:TB_system}). To calculate the necessary projected Green's function $G^P(z)$,  
we use  the Green's function technique of  Refs.~\cite{sanvito1999,sanvito2008}. 
Namely, we account for the effect of the leads attached to the normal region by means of 
the Dyson's equation:\cite{Dyson}
\begin{equation}
 G^P(z) = \left(z - H_{\rm eff}(z) - \Sigma_L - \Sigma_R \right)^{-1}\;. 
 \label{eq:G_Dyson}
\end{equation}
Here $\Sigma_L$ and $\Sigma_R$ are the self-energies of the left and right leads, respectively 
(see Section \ref{subsec:selfies} for further details).
$H_{\rm eff}(z)$ is the effective energy dependent Hamiltonian describing the surface of the normal 
region (see Fig.\ref{fig:TB_system}).
(The energy dependence of $H_{\rm eff}(z)$ can be interpreted as the effect of the inner sites located 
between the surface slabs $1$ and $N$.)
The effective Hamiltonian $H_{\rm eff}(z)$ can be obtained via several methods. 
For example, one can eliminate all the sites inside the central region by the decimation method and keep 
only the sites of the surface slabs \cite{smeagol-paper}.   
However, for long ballistic structures there is a more efficient method 
which we will briefly describe in Section \ref{subsec:Heff}.


\subsection{SNS system}
\label{subsec:sns-system}

The discussion in Sections~\ref{subsec:general-setup} and \ref{subsec:expct-current-op} 
was  general and would apply regardless of whether one assumes  normal or superconducting leads. 
In this section we discuss those aspects of the problem which are specific to 
normal-superconductor-normal (SNS) systems, i.e., 
systems where there are two superconducting terminals  and a central normal scattering region. 

We describe this inhomogeneous superconducting system by the Bogoliubov de Gennes (BdG) model.
Consequently, the Hilbert space of the superconducting system is constructed as the product of 
the Hilbert space of the normal system and the Nambu space describing the electron ($u$) 
end hole like ($v$) degrees of freedom. 
The matrix elements of $\hat{H}$ in Eq. (\ref{eq:fullHam}) can be written as
\begin{equation}
 H_0 = \left(
        \begin{array}{cc}
          H^{u}_0-\mu & \Delta \\
	  \Delta^{*} & H^{v}_0 + \mu
         \end{array}
          \right),
\end{equation}
and
\begin{equation}
 H_{\pm1} = \left( 
          \begin{array}{cc}
          H^{u}_{\pm1} & 0 \\
	  0 & H^{v}_{\pm1}
         \end{array}
         \right),
\end{equation}
where, for simplicity, we omitted the indexes labeling the slabs of the system.
Matrices $H^{u}_0$ and $H^{u}_{\pm1}$ describe the electron likes components.
The Hamiltonians of the hole like components are given by 
$H^{v}_0=-\left(H^{u}_0\right)^*$ and $H^{v}_{\pm1}=-\left(H^{u}_{\pm1}\right)^*$.
Finally, the diagonal matrix $\Delta$ contains the superconducting pair-potentials 
on the atomic sites of the slabs, and $\mu$ is the chemical potential.
Since in the central region the superconducting pair potential is zero, the electron and hole like 
components of the BdG equations become uncoupled and $H_{\rm eff}$ {becomes diagonal in the Nambu space}.
Similarly, if one calculates the current in the central (normal region), the current operator 
can be written in a block-diagonal form
\begin{equation}
 ^{S}\hat{I} = \left(
           \begin{array}{cc}         
           \hat{I}^{u} & 0 \\
	  0  & \hat{I}^{v}
          \end{array}
           \right)\;,
\end{equation}
where $\hat{I}_{u}$ and $\hat{I}_{v}$ are the charge current operators of the electron and hole like states, 
respectively (their explicit form for our  TB model is given in Section \ref{subsec:eff-current-op} ).  
The projected Green's function (see Eq.~\ref{eq:G_Dyson}), on the other hand,  has a matrix form 
\begin{equation}
 ^{S}G^P(z) = \left(
          \begin{array}{cc}           
           G^P_{uu} & G^P_{uv} \\
	  G^P_{vu}  & G^P_{vv}
          \end{array}\right)\;. 
          \label{eq:GP:BdG}
\end{equation}
where $G^P_{uv}$ and $G^P_{vu}$ are non-zero due to the fact that electron and hole components are 
coupled in the self-energies of the superconducting leads. 

Taking into account the thermal occupation of the 
electron-like  and hole-like states,  
we obtain the following expression for the charge current: 
\begin{eqnarray}
 I_{c,tot} &= \sum\limits_{E_{min}<E_n<E_{max}}\langle {}^{S}\hat{I}\rangle_{n,T}   \nonumber\\
      & =\frac{1}{\pi} {\rm Im} \int\limits_{\Gamma}dz\; f(z){\rm Tr}\left( G^P_{uu}(z)\hat{I}^{u}\right) \nonumber\\
 & +\frac{1}{\pi} {\rm Im} \int\limits_{\Gamma}dz\; \left(1-f(z)\right){\rm Tr}\left( G^P_{vv}(z)\hat{I}^{v}\right).
\label{eq:sumA_contour_fermi_S} 
\end{eqnarray}

\begin{figure}[htb]
\centering
 \includegraphics[width=6cm]{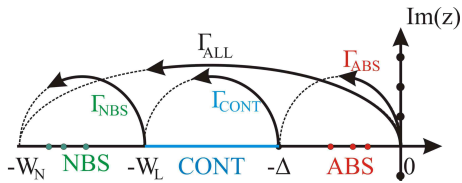}
 \caption{The integration paths to calculate the expectation values of the current operator 
 on the ABS, ScS and NBS states or to calculate the sum of the expectation values on all the 
  states at once. The singular points related to the Fermi distribution function that should 
  be  avoided by the integration contours
  are also indicated.
 \label{fig:contours}}
\end{figure}

In general, the spectrum of the BdG Hamiltonian is symmetrical around 
$E=0$ and therefore it is enough to consider 
either $E>0$ or $E<0$. Considering only the negative energies, 
the spectrum of the SNS junction can be divided into three spectral regions \cite{cini}.
The first region corresponds to the states of energy $-\Delta_0<E_n<0$. 
Due to the energy gap in the superconducting leads, these states are bounded to the normal region since 
they are decaying exponentially in the superconducting leads.
We refer to these states as the Andreev bound states (ABS).
The next energy regime is given by $-|BW_L| < E_n < -\Delta_0$, where $BW_L$ is the bandwidth 
of the superconducting leads, i.e., the maximal energy of the propagating states in the leads.
These scattering states form a continuous energy range in the spectrum.
Finally, in the case when the bandwidth of the normal region $BW_N$ is larger than $BW_L$, 
we can define a third energy regime.
Namely, for energies $-|BW_N| < E_n < -|BW_L|$ the states formed in the system are 
decaying exponentially in the leads, but are still propagating in the normal region.
According to Ref. \cite{cini}, we refer to these states as the normal bound states (NBS).
Figure \ref{fig:contours} shows the integration contours to be used to calculate the DC Josephson 
current due to  the ScS's, the ABS's and the NBS's, respectively. 
In addition, one can also calculate the contribution from all these states at once by integration 
over the contour $\Gamma_{ALL}$ also shown in Figure~\ref{fig:contours}.
{Note, that using this approach it is not necessary to find the zeros of a polynomial,
as in Reference~\onlinecite{cini}}. 

Finally, due to the fact that in superconducting systems the spectral density distribution is symmetrical 
with respect to $E=0$, the contour integration (\ref{eq:sumA_contour_fermi_S}) needs to be evaluated only 
in the ${\rm Re}(z)<0$ half-plane.
The contribution of the states in the ${\rm Re}(z)>0$ half-plane can be accounted for 
by a factor of $2$ in the final result.

This completes the general discussion of the TB method that we used to calculate the Josephson current. 
In the next section we discuss certain  model specific aspects of our calculations.


\section{Details of the TB calculations}
\label{sec:TB-details}

In this section we give the details of our TB calculations for SGS junctions 
which are relevant for obtaining  the results   presented in 
Sections \ref{sec:zero-field results} and \ref{sec:magnetic-field-osc}.

\subsection{Calculation of  the effective Hamiltonian} 
\label{subsec:Heff}

In this section we provide an {efficient} numerical method to calculate the effective Hamiltonian 
$H_{\rm eff}(z)$ needed to evaluate the expectation value of the current operator 
in Eqs. (\ref{eq:sumA_contour_fermi}) and (\ref{eq:sumA_contour_fermi_S}). 
While the procedure described here is optimized 
for a ballistic scattering region containing identical slabs, it has been shown 
in Ref.~\cite{graphene_antidot_sajat} that this approach can be generalized also to more 
complex geometries {as long as} the system is ballistic and is  numerically more  
efficient than the standard recursive Green's function techniques.

We assume that the central region is made of identical slabs described by the Hamiltonian 
$H_0^{i}\equiv H_0$ and coupled to each other by 
$H_{\pm1}^{i\mp 1,i}\equiv H_{\pm1}$ ($1\leq i \leq N$).
Following the procedure described in Ref. \cite{graphene_antidot_sajat}, we can obtain the $H_{\rm eff}(z)$ 
using the Green's function  $g_{i,j}$  of an infinite ribbon made of these slabs (here $i$, $j$ are slab indices). 
The Green's function  $g_{i,j}$ can be efficiently calculated using a semi-analytical 
formula introduced in Ref.~\cite{sanvito1999}. 
{In order obtain the elements of $H_{\rm eff}(z)$ we need to calculate} the propagators $g_{i,j}$ 
on slabs $i\in\{0,1,N,N+1\}$ and between these slabs.  
They can be arranged into a matrix that reads
\begin{equation}
 G(z) = \left(
       \begin{array}{cccc}
      g_{00}     & g_{01}      & g_{0N}     & g_{0(N+1)} \\
      g_{10}     & g_{11}      & g_{1N}     & g_{1(N+1)} \\
      g_{N0}     & g_{N1}      & g_{NN}     & g_{N(N+1)} \\
      g_{(N+1)0} & g_{(N+1)1} & g_{(N+1)N} & g_{(N+1)(N+1)}
     \end{array}\right).
\end{equation}
Since the structure of the ribbon contains only nearest neighbor couplings between the slabs without 
long range interaction, the effective Hamiltonian defined as $H(z) = zI -G(z)^{-1}$ has the following structure:
\begin{equation}
 H(z) = \left(\begin{array}{cccc}
      H_{00}     & H_{01}      & 0          & 0 \\
      H_{10}     & H_{11}      & H_{1N}     & 0 \\
      0          & H_{N1}      & H_{NN}     & H_{N(N+1)} \\
      0          & 0            & H_{(N+1)N} & H_{(N+1)(N+1)}
     \end{array}\right),
\end{equation}
Note that there is no coupling between the slabs $0$ and $N$ since these slabs are coupled via the slab $1$, 
and therefore the matrix element $H_{0N}$ vanishes.
For similar reasons the matrix elements $H_{N0}$, $H_{1(N+1)}$, $H_{N0}$, $H_{(N+1)0}$ and $H_{(N+1)1}$ also become zeros.
Let us now apply a perturbation to the Hamiltonian $H$ given by
$V_1 = -H_{01}|0\rangle\langle 1| - H_{10}|1\rangle\langle 0|$ and 
$V_2 = -H_{N(N+1)}|N\rangle\langle N+1| - H_{(N+1)N}|N+1\rangle\langle N|$, where $|i\rangle$ 
represents the subspace of the $i$th slab.
The potentials $V_1$ and $V_2$ uncouple the scattering region containing of $N$ slabs 
from the rest of the infinite ribbon. 
Therefore the inner $2\times2$ part of the Hamiltonian $H_{z}+V_1+V_2$ describes the 
the effective Hamiltonian  of the central region, which can be written 
in the following form: 
\begin{equation}
 H_{\rm eff}(z) = \left(\begin{array}{cc}
             H_{11}(z)      & H_{1N}(z) \\
	     H_{N1}(z)      & H_{NN}(z)
            \end{array}\right). \label{eq:Heff}
\end{equation}
The effect of the sites between the surface slabs are incorporated within the energy dependence of $H_{\rm eff}$.
We note  that using the decimation method\cite{smeagol-paper} one would obtain an identical 
effective Hamiltonian. However, the described procedure involves only sites that are located 
on the surface of the central region and therefore the computational cost of calculating  
$H_{\rm eff}$ is scaling only with the width of the ribbon. 
This is especially important in the case of \textit{long} scattering regions.


\subsection{The effective current-operator}
\label{subsec:eff-current-op}

In order to evaluate the contour integral given in (\ref{eq:sumA_contour_fermi_S}) one needs to construct 
the matrix representation of the current operator between the surface slabs of the central region.
Since the electron and hole like components are uncoupled in the 
central region, one can obtain individual charge current operators for the electron ($\hat{I}_{u}$) 
and hole ($\hat{I}_{v}$) like components.
Each of them can be derived, similarly to the normal systems, from the 
{corresponding} effective Hamiltonian $H_{\rm eff}$ 
by means of a discretized continuity equation. 
Thus, we obtain the current operator between 
the slabs $1$ and $N$ from the elements of the effective Hamiltonian (\ref{eq:Heff}).
Making use of the block diagonal structure of $H_{\rm eff}$, the current operator reads
\begin{equation}
 \hat{I}^{u/v}(z) = \frac{e}{\hbar}
                     \left(
                     \begin{array}{cc}                    
                     0 & -H^{u/v}_{1N}(z) \\
		  -H^{u/v}_{N1}(z) & 0
                    \end{array}\right).
\end{equation}
Similarly to the effective Hamiltonian, the effective current operator is also energy dependent.


\subsection{Implementing the magnetic field in the SNS junction}
\label{subsec:Bfield}

We now discuss how the  magnetic field  is taken into account in our calculations. This is needed 
in order to calculate the oscillations of the critical current, see Section~\ref{sec:magnetic-field-osc}.
We limit our considerations to low magnetic fields where the effect of screening currents on $\Delta$ 
can be neglected. Consequently,  the only effect of the induced supercurrents 
on the surface of the superconducting regions is that the magnetic field is expelled 
from the superconducting leads, but is considered to be homogeneous in the normal region.
\begin{figure}[hbt]
\centering
\includegraphics[scale=0.65]{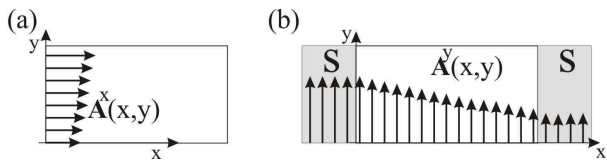}
 \caption{(a) The vector potential $\mathbf{A}^x(x,y)$ in the normal region describing a homogeneous 
 magnetic field $B_z$ perpendicular to the graphene sheet.
(b) The vector potential $\mathbf{A}^y(x,y)$ in the whole SNS junction after the gauge transformation. 
In the superconducting contacts the magnetic field is zero, while in the normal region the 
strength of the magnetic field is $B_z$ (see the text for details).
\label{fig:vector_potential}}
\end{figure}
The corresponding vector potential can be given in a Landau gauge
\begin{equation}
 \mathbf{A}^y(x,y) = \left(\begin{array}{cc}
                      0 \\ A_y(x)
                     \end{array}\right),
\end{equation}
where
\begin{equation}
  A_y(x) = \left\{\begin{array}{ll}
                  A_0 & \textrm{if } x\leqslant 0 \\
		  A_0 + xB_z & \textrm{if } 0< x\leqslant L \\
		  A_0 + LB_z & \textrm{if } x > L
                 \end{array}\right.
\label{eq:Ay-Landau-g}                 
\end{equation}
As we have seen in  section \ref{subsec:Heff}, one can calculate  the 
effective Hamiltonian $H_{\rm eff}$  efficiently
for long ballistic systems if the normal region consists of identical slabs. This translational invariance
would be broken by the vector potential given in Eq.~(\ref{eq:Ay-Landau-g}). 
To avoid this problem  one may try to  incorporate the magnetic field in the normal region by a vector 
potential $\mathbf{A}^x$  that is translational invariant in the normal region 
[see Fig. \ref{fig:vector_potential}.(a)]:
\begin{equation}
 \mathbf{A}^x(x,y) = \left(\begin{array}{c}
                      yB_z \\ 0
                     \end{array}\right).
\end{equation}
Note however that  $\mathbf{A}^x$ cannot be fitted continuously to the 
constant vector potential in the superconducting leads. 
However,  $\mathbf{A}^y(x,y)$ is related to $\mathbf{A}^x(x,y)$  by a gauge transformation 
$\mathbf{A}^y(x,y) = \mathbf{A}^x(x,y) + {\rm grad}\chi(x,y)$ with a gauge field given as
\begin{equation}
 \chi(x,y) = \left\{\begin{array}{ll}
                  A_0x & \textrm{if } x\leqslant 0 \\
		  A_0x - xyB_z & \textrm{if } 0< x\leqslant L \\
		  A_0x + LyB_z & \textrm{if } L< x
                 \end{array}\right.
\label{eq:chi-phase}                 
\end{equation}
Since the magnetic field enters the calculations through the Peierls-substitution 
(see Eq.~\ref{eq:TB-peierls}), one can show that the effect of this  gauge transformation 
on the effective Hamiltonian 
$H_{\rm eff}$ can be expressed as
\begin{equation}
 \tilde{H}_{\rm eff} = UH_{\rm eff}U^{\dagger},
\end{equation}
where $U_{i,j} = \delta_{i,j} \exp[i \chi(\mathbf{R}_i)]$ is a matrix describing 
a unitary transformation. Here $\chi(x,y)$ is defined by Eq.~(\ref{eq:chi-phase}) and 
$\mathbf{R}_i$ is a lattice vector.


\subsection{Calculation of the self energies}
\label{subsec:selfies}

Finally, we briefly discuss the calculation of the self-energies $\Sigma_{L,R}$ that enter 
Eq.~(\ref{eq:G_Dyson}). 
We obtained them using the model of References~\onlinecite{Carlo_short1,Carlo_revmod}, i.e., 
assuming that the superconducting leads 
consists of highly doped semi-infinite graphene ribbons 
where a finite superconducting pair potential $\Delta$ was induced by proximity effect. 
The surface Green's functions of the leads and the corresponding self-energies $\Sigma_{L,R}$
can be calculated as described in Reference~\onlinecite{sanvito2008}.
We note that other approaches to calculate $\Sigma_{L,R}$, such as the 
``bulk-BCS'' model discussed in Reference~\onlinecite{martin-rodero},  
could equally be used in the computational framework we introduced.


\section{Summary and Outlook}
\label{sec:summary}

In summary, we have studied theoretically the DC Josephson current in long SGS junctions.
We developed a  theoretical framework that can be applied to an arbitrary 
superconducting-normal-superconducting junction defined on a tight-binding lattice. 
By treating the bound and scattering states on an equal footing it presents an  
accurate and efficient numerical method to calculate the equilibrium Josephson current.

We used this theoretical approach  to investigate the dependence of the critical current 
on the geometrical properties of the junctions and on the magnetic field 
in the ballistic transport regime. 
In the zero field and low temperature limit we have found that the critical current 
decays as $\sim\xi_0/L$ in agreement with recent measurements.
For temperatures comparable to $T_c$, on the other hand, the critical current becomes 
exponentially suppressed.
Furthermore, we have found that  in the long junction limit
the contribution of the ScS to the Josephson current is as important as are the 
contribution coming from the ABSs.

We have also studied the magnetic oscillations of the critical current. 
Generally, for a given magnetic field 
one can distinguish the high-doping and the low-doping limits which are 
defined in terms of the  cyclotron radius $R_c$ as  $R_c \gtrsim W,L$ and $R_c \lesssim W,L$, respectively. 
In the high-doping regime, when  $R_c$ is much larger than other length scales 
of the junction, the the period of the oscillations depends on 
the geometry of the junction in a similar way as predicted by previous 
quasiclassical calculations for S-2DEG-S Josephson junctions. 
For wide junctions, i.e. $W/L > 1$, we recover the Fraunhofer-like oscillation pattern, however, 
deviations from this oscillation pattern  start to appear at a magnetic flux $\Phi\sim W/L\;\phi_0$ 
resulting in an increased oscillation period.
For narrow junctions, in turn, one can observe a more complex interference pattern in 
the magnetic field dependence of the critical current without reaching the zero value.
Close to the Dirac point $R_c$ becomes comparable to the dimensions of the junctions, 
thus the dynamical effects, due to the  curved semiclassical trajectories, 
cannot be neglected any more.
According to our results, the interplay of these dynamical effects and the magnetic field
induced quantum mechanical interference has twofold effect on the critical current.
Firstly, by decreasing $R_c$ the oscillation period starts to increase and the 
minimums of the critical current are lifted from zero.
Secondly, for $R_c$ smaller than the length of the junction, the magnetic dependence of 
the critical current does not show any regular oscillations.
We note, however, that in this work we did not address  the case  $2 R_c < L, W$, i.e., when 
the formation of  quantum Hall states is expected. 

{The methodology that we have introduced here is quite flexible and  
it would allow to address a number of further problems. As a brief outlook, we would mention 
the following ones. Firstly, the interplay of Landau quantization and Josephson current flow in a 
weak link, i.e., the regime $2 R_c < L, W$. This question is timely, as the first report on the observation 
of supercurrent in the quantum Hall regime in a SGS junction  has recently appeared \cite{finkelstein}.
Secondly, although we have focused on ballistic junctions in this work, disorder effects 
in the normal region can also be incorporated using, e.g., the recursive Green's function 
technique. We mention two problems in this regard: in  Reference~\onlinecite{yacobi2015a} 
an anisotropic supercurrent distribution was found at low
dopings, where the role of disorder is expected to be especially important. 
This was explained by calculating the normal density of states and assuming that it also 
determines the supercurrent flow. One could verify this assumption by solving the  
BdG equations for a disordered normal region and calculating the supercurrent distribution.
Finally, we note that Josephson vortices were predicted\cite{cuevas, bergeret} and later 
experimentally verified\cite{roditchev} to exist in diffusive SNS junctions. One  may expect that 
they also appear in diffusive SGS junctions. 
}


\section{Acknowledgement}
 
A. K. acknowledges discussions with Srijit Goswami about their results 
in Reference \onlinecite{vandersypen}. 
P. R. and J. Cs. acknowledges the support of the OTKA through the grant 
K108676. 
A. K. was   supported  by the Deutsche Forschungsgemeinschaft (DFG) 
through SFB767. 
P. R. acknowledges the support of the MTA postdoctoral research program 2015.




\begin{thebibliography}{99}
\bibitem{calcs-detail}
Our calculations are performed, { if not indicated otherwise,}
on  {zigzag} nanoribbons of  { fixed} widths $W$. 
To change the aspect ratio $W/L$ we change the length $L$ of the junction.
We also choose  to keep the ratio $L/\xi_0$ fixed, which means that for 
a given chemical potential $\mu_N$ the ratio $\mu_N/\Delta_0$ also changes.
\end{thebibliography}
\end{document}